\documentclass[twocolumn]{aastex631}

\usepackage{multirow}
\usepackage{amsmath}

\shorttitle{Paper I}
\shortauthors{Balakrishnan et al.}

\newcommand{\Chandra}{{\sl Chandra}}

\newcommand{\SgrA}{Sgr~A*}
\newcommand{\FeKalpha}{Fe~K$\alpha$}

\begin{document}

\title{Multistructured accretion flow of Sgr A* I: Examination of a RIAF model}

\correspondingauthor{Mayura Balakrishnan}
\email{bmayura@umich.edu}

\author[0000-0001-9641-6550]{Mayura Balakrishnan}
\affil{Department of Astronomy, University of Michigan, 1085 S. University \\
Ann Arbor, MI 48109, USA}

\author[0000-0002-5466-3817]{Lia Corrales}
\affil{Department of Astronomy, University of Michigan, 1085 S. University \\
Ann Arbor, MI 48109, USA}

\author[0000-0001-9564-0876]{Sera Markoff}
\affiliation{Anton Pannekoek Institute for Astronomy/GRAPPA, University of Amsterdam, Science Park 904,
1098 XH Amsterdam, Netherlands}

\author[0000-0001-6923-1315]{Michael Nowak}
\affiliation{Department of Physics, Washington University at St. Louis, 1 Brookings Dr, St. Louis, 63130 MO, USA}

\author[0000-0001-6803-2138]{Daryl Haggard}
\affiliation{Department of Physics, McGill University, 3550 University Street \#040, Montreal, QC H3A 2A7, Canada}

\author[0000-0002-9279-4041]{Q.\ Daniel Wang}
\affiliation{Department of Astronomy, University of Massachussets Amherst, 710 North Pleasant Street Amherst, MA 01003, USA}

\author[0000-0002-8247-786X]{Joey Neilsen}
\affiliation{Department of Physics, Villanova University, 800 Lancaster Avenue, Villanova, PA 19085, USA}

\author[0000-0002-9213-0763]{Christopher M.\ P.\ Russell}
\affiliation{Department of Physics and Astronomy, Bartol Research Institute, University of Delaware, Newark, DE 19716, USA}

\author[0000-0002-9019-9951]{Diego Calder\'on}
\affiliation{Hamburger Sternwarte, Universität Hamburg, Gojenbergsweg 112, D-21029 Hamburg, Germany}

\author[0000-0003-1965-3346]{Jorge Cuadra}
\affiliation{Faculty of Liberal Arts, Universidad Adolfo Ibañez, Av. Padre Hurtado 750, Viña del Mar, Chile}
\affiliation{Millennium Nucleus on Transversal Research and Technology to Explore Supermassive Black Holes (TITANS)}

\author[0000-0003-3852-6545]{Frederick Baganoff}
\affiliation{MIT Kavli Institute for Astrophysics and Space Science, MIT, 70 Vassar St, Cambridge, MA 02139, USA}



\begin{abstract}

The extreme low-luminosity supermassive black hole \SgrA\ provides a unique laboratory in which to test radiatively inefficient accretion flow (RIAF) models. Previous fits to the quiescent \Chandra\ ACIS-S spectrum found a RIAF model with an equal inflow-outflow balance works well. In this work, we apply the RIAF model to the \Chandra\ HETG-S spectrum obtained through the \Chandra\ X-ray Visionary Program, which displays features suggestive of temperature and velocity structures within the plasma.  A comprehensive forward model analysis accounting for the accretion flow geometry and HETG-S instrumental effects is required for a full interpretation of the quiescent \Chandra\ HETG-S spectrum. We present a RIAF model that takes these effects into account. Our fits to the high-resolution gratings spectrum indicate an inflow balanced by an outflow ($s \sim 1$) alongside a temperature profile that appears shallower than what would be expected from a gravitational potential following $1/r$. The data require that the abundance of Iron relative to solar is $Z_{Fe} < 0.32 Z_\odot$ (90\% credible interval), much lower than the $2~Z_\odot$ metallicity measured in nearby late-type giants. While future missions like \textit{NewAthena} will provide higher spectral resolution, source separation will continue to be a problem. Leveraging \Chandra's unparalleled spatial resolution, which is not expected to be surpassed for decades, remains essential for detailed investigations of the densely populated Galactic Center in X-rays.

\end{abstract}



\section{Introduction} \label{sec:intro}

Active Galactic Nuclei (AGN) feedback is an important mechanism by which star formation can be quenched \citep[see, e.g.][]{Weinberger2018, Vogelsberger2014, Fabian2012}, significantly shaping galaxy properties over cosmic time. However, even in low-luminosity states, mechanical feedback from radiatively inefficient supermassive black holes (SMBHs) can still have a significant effect on galactic evolution \citep{Yuan2018}. Low-luminosity AGNs (LLAGNs; $L \leq 10^{39} - 10^{40} {\rm erg/s}$) constitute the majority of observed AGN in the nearby universe \citep{Ho2008}. The closest supermassive black holes, detectable by modern instruments such as Keck, Hubble, EHT, and \Chandra, are M31* \citep[$L \sim 10^{38}$ erg/s; $M_* \sim 10^8 M_\odot$;][]{Bender2005} and our own Milky Way's Sagittarius A* \citep[\SgrA; $L_{X} \sim 10^{33}$ erg/s; $M_* = 4.1 \times 10^6 M_\odot$;][]{Baganoff2003}. Both are categorized as low-luminosity active galactic nuclei (LLAGN). The proximity of \SgrA\ allows for in-depth investigations into the dynamics between supermassive black holes and their surroundings, deepening our understanding of LLAGNs.

\SgrA\ displays many behaviours that make it a dynamic astrophysical laboratory. The SMBH has significant time variability, with flares that reach 100~-~1000 times the quiescent luminosity in both the X-ray  \citep[on average 1 per day;][]{Nowak2012, Neilsen2013, Haggard2019} and NIR \citep[4 per day;][]{Schodel2011}. During non-flaring periods, the SMBH maintains a low luminosity of approximately $L_X \sim 10^{33}~\rm erg \ s^{-1}$  = $10^{-10} L_{\rm edd}$ \citep{Baganoff2003,Wang2013,Corrales2020}, despite the material available at the Bondi radius ($\sim 4 - 5 \arcsec$; see Section \ref{sec:model_params}). Around 30 massive Wolf-Rayet (WR) stars orbit within 0.4 pc ($\sim$ 10\arcsec) with mass loss rates of $\sim 10^{-5}-10^{-6} M_\odot\ \rm yr^{-1}$ \citep{Martins2007,Cuadra2015}, providing plasma for \SgrA\ to accrete. Meanwhile, Faraday rotation measurements and Event Horizon Telescope (EHT) models closer to the horizon ($\sim 4 \times 10^{-6}$ pc) show a mass accretion rate closer to $10^{-9}-10^{-8} M_\odot\ \rm yr^{-1}$ \citep{Marrone2007, EHT_PaperI, EHT_PaperVIII}. 
Advection-dominated accretion flows \citep[ADAFs;][]{Narayan1994,Narayan2008}, in which a significant portion of the accretion energy is retained in ions and electromagnetic radiation losses are minimal, were developed to explain the observed low-luminosity of \SgrA. However, measurements of the low accretion rate (in addition to issues with self-similarity in the ADAF solution) suggest that advection-dominated inflow-outflow solutions \citep[ADIOS;][]{Blandford1999, Blandford2004}, where some of the accreted material is expelled in an outflow, may better describe the physical processes in the \SgrA\ accretion flow.
Magnetic fields have also been theorized to prevent matter from flowing into the inner regions \citep{pang2011phd}, but their importance is unclear; modelling of EHT data finds that magnetic fields are required, but some general relativistic magnetohydrodynamic (GRMHD) simulations do not support this scenario \citep{ressler2020}.

Regardless of underlying physics, optically thin, quasi-spherical accretion flows with a variety of internal conditions can be parameterized with the Radiatively Inefficient Accretion Flow (RIAF) prescription \citep{Quataert2002, Yuan2003, Yuan2014}. RIAFs typically adopt a simple power-law to describe the mass accretion rate, $\dot{m} \propto r^s$, and material density, $n \propto r^{-3/2+s}$, where $s = 0$ corresponds to classical Bondi accretion and $s \sim 1$ indicates an inflow balanced by an outflow \citep{Begelman2012}. Observations of the SMBH at different wavelengths probe different scales: the event horizon can be seen through long baseline radio interferometry, radio polarization probes within $\sim$ 10 $R_g$ (where $R_g = GM_*/c^2$), and the quiescent X-ray emission originates from $10^4 - 10^6 R_g$, around the Bondi radius. Consequently, the multiwavelength spectral energy distribution (SED) offers comprehensive insight into the underlying physics, potentially exhibiting different $s$ values at different radii. One model that successfully explains the multiwavelength SED of \SgrA\ and can account for the observed flaring behavior is \citet{Yuan2003}, where thermal electrons with $s \sim 0.3$ in collisional ionization equilibrium (CIE) transfer $\sim 1.5$\% of their energy to a population of non-thermal electrons. Newer fits incorporating different boundary conditions have found $s \sim 0.05$ within 30$R_g$ and $s \sim 0.6$ closer to $R_{Bondi}$ \citep{Ma2019}, indicating that the outflow increasingly dominates as distance from the SMBH increases. Simulations confirm this, with many requiring some sort of outflow that dominates outside $R > 100 R_g$ \citep{Yuan2014,Ressler2018,Dexter2020,Chatterjee2021}.

To investigate the X-ray properties of \SgrA, the \Chandra\ X-ray Visionary Program (PIs: Markoff, Nowak, \& Baganoff)\footnote{https://www.sgra-star.com/} obtained 3~Ms of \Chandra\ exposures in 2012, utilizing the \Chandra\ High Energy Transmissions Grating Spectrograph \citep[HETG-S;][]{CanizaresHETG}. Other works focus on deciphering the time variability of \SgrA, but here we summarize results from previous analysis of the quiescent emission. \citet{Wang2013} (hereafter W13) analyzed the zeroth order (non-dispersed) quiescent spectrum and found emission lines corresponding to a plasma with $T \sim 3.5$ keV and a RIAF $s \sim 1$. The plasma being accreted onto the SMBH primarily originates from shocked, colliding stellar winds from the 30 WR stars in the central parsec. \citet{Roberts2017} performed hierarchical Bayesian fits, with initial conditions matching the WR orbits, to the 2D image of \SgrA\ with hydrodynamic models. They found that 99\% of the accreted mass is ejected in a polar outflow before reaching the horizon. Smoothed particle hydrodynamics (SPH) simulations of these stellar winds are also able to recreate the observed X-ray spectrum \citep{Russell2017}. We address whether these simulations are better able to reproduce the observed HETG-S spectrum in our companion paper, Balakrishnan et al. 2024b, hereafter referred to as Paper II.

Studying the central $\sim$ 5$-$10 arcseconds ($1 - 2 \times 10^6 R_g$) around \SgrA\ in X-rays poses a challenge due to high extinction and the presence of nearby contaminating objects. Extracting the high-resolution gratings spectrum requires meticulous management of flares, point sources, and nearby contaminants. Although use of \Chandra's gratings allows for increased spectral resolution compared to the CCD, the dispersed light from the slit-less gratings intertwines with emissions across the entire field-of-view, which is not a problem for bright point sources. Meanwhile, \SgrA\ exhibits X-ray brightness comparable to that of the extended and diffuse pulsar wind nebula (PWN) G359.95-0.04 and the massive star cluster IRS 13E, both of which lie an angular projected distance of 4\arcsec\ (only 8 \Chandra\ pixels) from the SMBH and appear to extend into the Bondi radius of the accretion flow. Using custom background regions and non-standard extraction techniques to obtain the high-resolution spectrum, \citet{Corrales2020} identified various line features within the \FeKalpha\ complex potentially associated with Fe XXVI, Fe XXV, Fe XX, Fe XXII, and Fe XXIII. The detection of multiple ionization states of Iron suggests the possibility of probing multiple temperature layers within the accretion flow. Moreover, apparent line shifts suggest the potential for investigating the velocity structure and orientation of the accretion flow using the high-resolution spectrum. However, this line structure could be a stochastic artifact, as each spectral bin only contains a handful of counts. Therefore, a comprehensive forward model analysis encompassing the geometry, instrumental effects, and anticipated velocity structure within the \SgrA\ accretion flow is imperative for a complete interpretation of the \Chandra\ HETG-S spectrum.

The goal of this work is to fit the \Chandra\ HEG spectrum with a RIAF model to investigate whether the apparent lower energy line features can be explained by temperature and density structure in the \SgrA\ accretion flow. We build on the previous work of \citet{Corrales2020}, fitting the high-resolution quiescent X-ray spectrum with careful consideration of the system geometry, \Chandra\ optics, and the HETG order-sorting algorithm. Focusing on the \FeKalpha\ complex, the strongest line features found in the HEG spectrum \citep{Corrales2020}, we fit the 6.4 - 7.2~keV spectrum with a RIAF model to constrain the temperature and density profiles of gas about $10^4 R_g - 10^6 R_g$ from the SMBH. In Section 2, we introduce the unique challenges presented by the \Chandra\ HETG instrument and the high-resolution spectrum used in this work. Section 3 outlines the procedure used to calculate the plasma emission and describes how we incorporated the unique geometry and spectral extraction techniques into our forward model.  In Section 4, we fit the data and discuss our interpretations, and end with conclusions. In Paper II, we extend our analysis by fitting the data to synthetic spectra built on hydrodynamic simulations based on \citet{Cuadra2015} and \citet{Russell2017} of the nearby WR stellar winds, and compare the two models.

\section{Dataset \& Chandra Instrumentation} \label{sec:data}

\begin{figure*}
    \centering
    \includegraphics[width=0.491\textwidth]{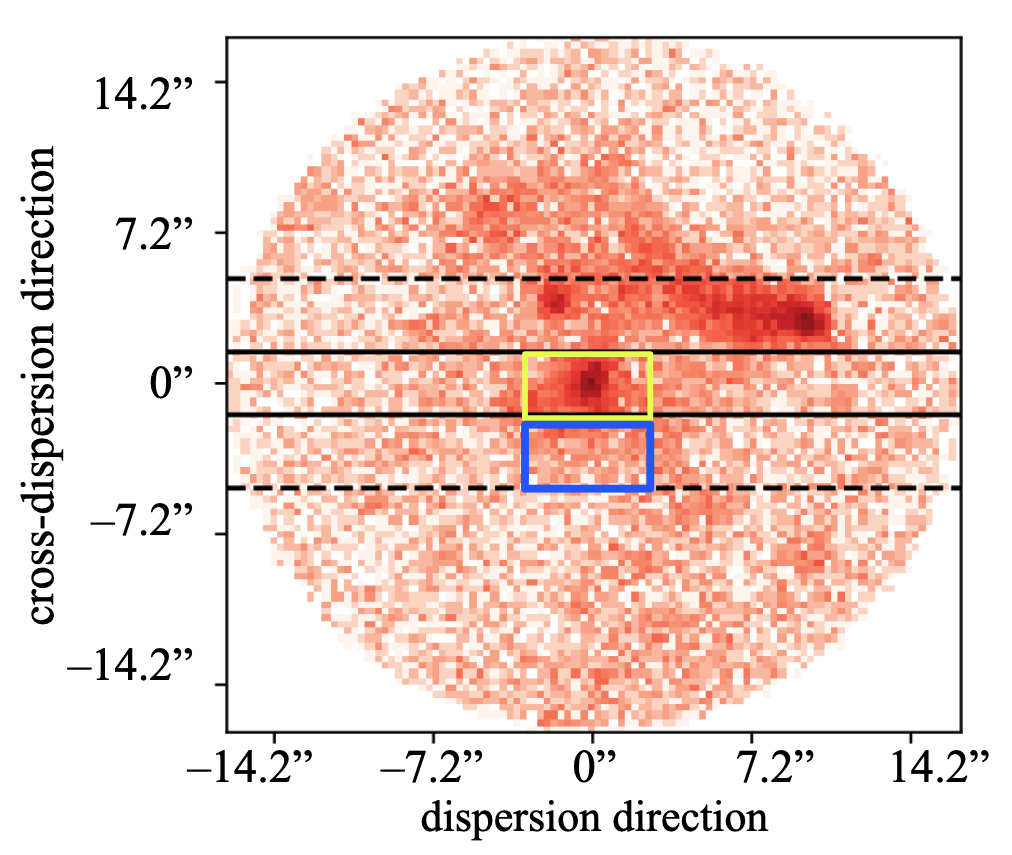}
    \includegraphics[width=0.502\textwidth]{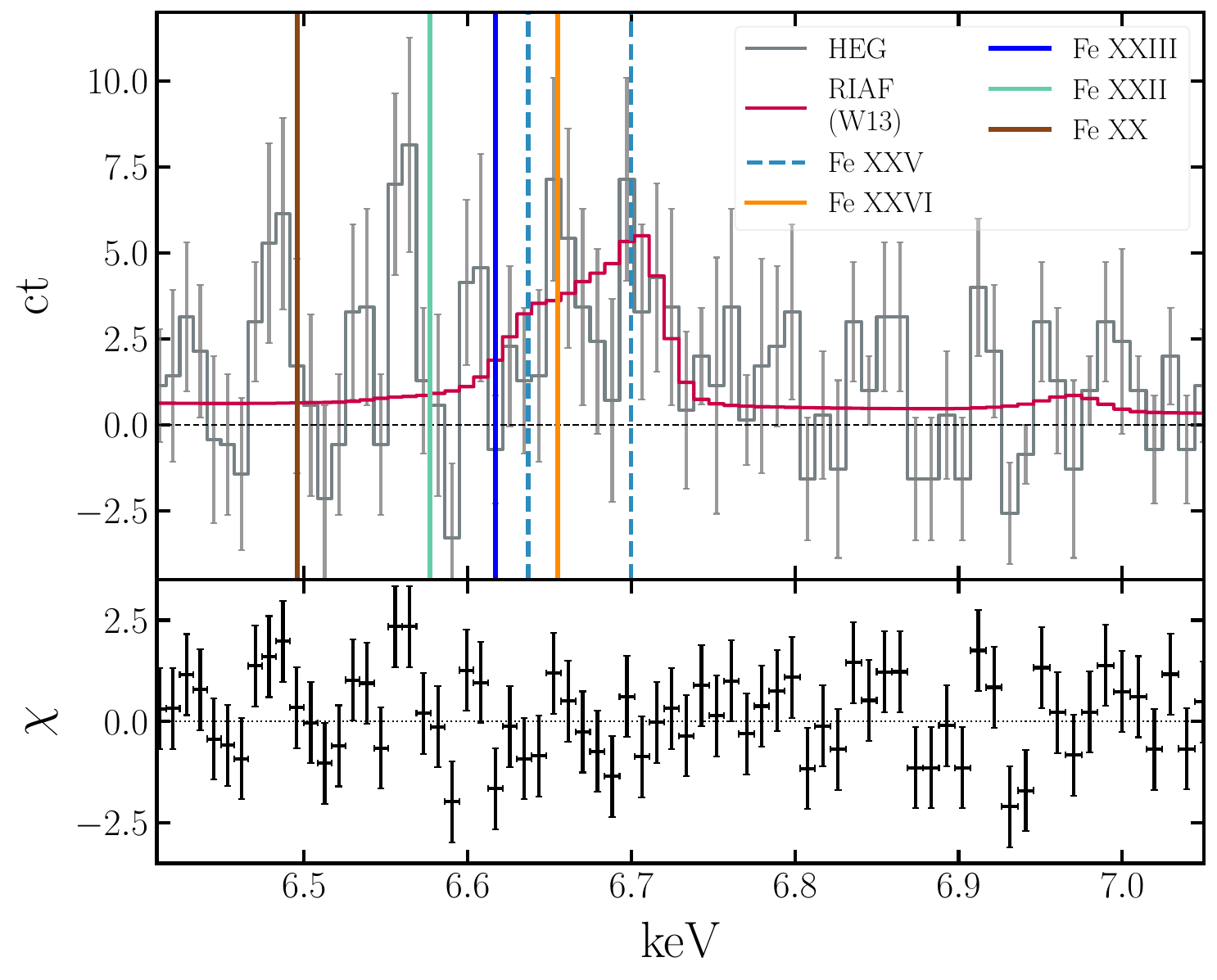}
    \caption{LEFT: Adaptation of Figure 1 from \citet{Corrales2020} showing the cleaned, quiescent 2 - 8 keV data for one roll angle. The 0th order (non-dispersed) image is displayed in HEG coordinates with the dispersion direction on the x-axis and the cross-dispersion direction on the y-axis. Solid black lines bracket the source extraction region out to 1.5\arcsec, while the dashed lines denote the background extraction region, 5\arcsec~away from \SgrA\ in the cross-dispersion direction. The yellow and blue boxes refer to the effective regions of the image that contribute to the source and background spectra, respectively. RIGHT: Adapted from Figure 6 of \citet{Corrales2020}, showing the the combined, unbinned, stacked and background subtracted HEG $\pm 1$ counts histogram in black. The best-fit RIAF model from W13 is plotted in red, showing that the background subtracted spectrum has additional structure. The Fe XXV doublet wavelengths are plotted with blue dashed lines, and Fe XX, Fe XXII, Fe XXII, and Fe XXVI are plotted in brown, teal, blue, and orange lines respectively.}
    \label{fig:data}
\end{figure*}

\Chandra\ contains two focal plane science instruments: the Advanced CCD Imaging Spectrometer (ACIS) and the High Resolution Camera, along with two gratings spectrometers for high and low energies. The High Energy Transmission Grating Spectrometer \citep[HETGS;][]{CanizaresHETG} delivers a spectral resolution of 60-1000 in the 0.4-10 keV range and disperses light across six ACIS-S CCD chips. The HETG itself disperses light along two axes, the Medium Energy Grating (MEG), sensitive between 0.4 and 5 keV, and the High Energy Grating (HEG), capturing light between 0.8 and 10 keV. 

The HEG has significant sensitivity in its operational range, with a resolution of $E/dE \sim 170$ at 6.7 keV for point sources. However, the resolution is notably impacted by the extended and diffuse nature of \SgrA. The HETG functions as a slitless spectrograph; dispersed light from the gratings forms a characteristic `X' pattern, with photons landing directly onto the ACIS-S detector. Each photon is scattered in the dispersion direction corresponding to the energy of the photon. An order-sorting algorithm decides whether or not a photon at a particular dispersion angle came from the central source based on its CCD estimated energy ($\Delta E \sim 100$ eV), and assigns photons to different diffraction orders by the use of ``banana plots'' (Section 8.2 of \Chandra\ Proposal Observatory Guide; CPOG \footnote{https://cxc.harvard.edu/proposer/POG/html/chap8.html}). The extended nature of the \SgrA\ image poses challenges for effectively and precisely isolating photons originating from the accretion flow amidst nearby contaminating objects or background sources. The extraction pipeline may therefore inadvertently capture contaminating photons that do not stem from the accretion flow.

The HEG~+1 and HEG~-1 orders (opposite sides of the `X') exhibit several distinctions. Photons with energies $\sim$ \FeKalpha\ dispersed in the $m = -1$ direction primarily land on ACIS chip S1 (back-illuminated), while those in the $m = +1$ direction land on chips S4/S5 (front-illuminated). This  results in differences in effective area and line response functions for both orders. While these effects are minimal for bright sources, they become more pronounced in low-count scenarios, necessitating separate analysis and fitting of the HEG~+1 and HEG~-1 order data in our work.

The dataset used in this work was carefully extracted \citep{Corrales2020} from the 3~Ms \Chandra\ X-ray Visionary Program (PIs: Markoff, Nowak \& Baganoff); in this paper we fit this high-resolution spectrum of \SgrA \ in quiescence with a RIAF model. \citet{Corrales2020} removed background and Sgr A* flares, resulting in 2.55 Ms total exposure. Observations were stacked by roll angle, and background spectra were generated by selecting regions ``up'' or ``down'' opposite to contaminating sources PWN G359.95-0.04 and IRS 13E along the cross-dispersion direction. The extracted source and background spectra encompass regions from 0\arcsec~-~1.5\arcsec, and 1.5\arcsec~-~5\arcsec, respectively. In the left panel of Figure \ref{fig:data}, we highlight the effective image regions probed by the HEG. Source and background regions are shown in yellow and blue, respectively. The width of the boxes are determined by the finite energy extraction width used in order sorting (via the \texttt{ciao} tool \texttt{tg\_resolve\_events}, more details in Section \ref{sec:hetg-geometry}).  The raw combined background-subtracted spectrum, depicted in the right panel of Fig~\ref{fig:data}, exhibits strong emission \FeKalpha\ emission from FeXX$-$XXVI features from the inner region of the accretion flow. The HEG spectrum is overplotted with the $s \sim 1$ RIAF from W13 in red, which best fits the zeroth order, low-resolution spectrum. In this analysis, we focus specifically on the \FeKalpha\ complex, the most prominent X-ray emission feature from \SgrA, and limit the analysis to 6.4 keV $-$ 7.2 keV. 

\section{Methods} \label{sec:methods}

\subsection{RIAF plasma model for the accretion flow}\label{sec:riaf}

To construct our model, we use the parametric RIAF interpretation \citep{Quataert2002, Yuan2003, Yuan2014}, where the electron temperature and density profiles are given as follows:
\begin{equation}\label{eq:RIAF}
    \begin{split}
        T(r) &= T_0 \left( \frac{r}{r_0} \right)^{-q} \\
        n_e(r) &= n_{e,0} \left( \frac{r}{r_0} \right)^{-3/2 + s}
    \end{split}
\end{equation}
We fix $T_0$ and $n_{e,0}$ to the temperature and density conditions reported in \citet{Baganoff2003}, and fix $M_{\rm SMBH}=4.15\times10^{6}M_\odot$ \citep{gravity2019}. The RIAF parameter $s$ describes the balance between outflow and inflow, with $s = 0$ corresponding to classical Bondi accretion and $s = 1 $ indicating an inflow equally balanced by an outflow \citep{Begelman2012}, while $q$ represents the steepness of the temperature profile. To calculate the expected emission arising from a RIAF, we approximate the accretion flow with shells of plasma assumed to be optically thin and in collisional ionization equilibrium (CIE). We use the publicly available \texttt{pyAtomDB}\footnote{https://github.com/atomdb/pyatomdb} version 0.10.13 code with AtomDB version 3.0.9 to model the photon emissivity in each spherical shell. The plasma is assumed to be optically thin and spherically symmetric, such that the total spectrum is obtained by numerically integrating the luminosity function over a series of nested shells with differential volume element $dV$. We calculate the luminosity at each radius using \texttt{pyatomdb} for a given RIAF $s$, $q$, and $Z$. 

The python module \texttt{pyAtomDB} can be used to calculate emission for plasmas up to $kT \approx 86$ keV, or $T \approx 10^9$ K.  Above this temperature ceiling, astrophysical plasmas are expected to be fully ionized, producing a smooth continuum with no atomic features. For plasma shells where the temperature exceeds $kT > 86$~keV, we calculate the relativistic bremsstrahlung continuum using the pyAtomDB functions \texttt{do\_brems} and \texttt{calc\_ee\_brems}, which calculate the ion-electron and electron-electron bremsstrahlung emissivity, respectively. Around $86$ keV, the inclusion of electron-electron bremsstrahlung adds about $20\%$ to the continuum emission. The point at which the accretion flow plasma reaches these temperatures depends on the slope of the RIAF temperature profile. For $q \sim 2$ and $q \sim 1$, the plasma becomes fully ionized for radii $R < 10^5 R_g$ and $R < 10^4 R_g$, respectively.

In reality, the accretion flow is not spherically symmetric. In a companion paper, Balakrishnan et al. 2024b, we compare our RIAF model to simulations of the WR stellar winds surrounding \SgrA. The two models offer different physical insights; the simulations take into account velocity broadening of the colliding winds and asymmetries in the plasma, while the spherically symmetric RIAF model incorporates an inflow/outflow balance but does not include velocity structure. We do not include thermal broadening in our RIAF model since its impact is negligible; thermal broadening in a $10^7 K$ plasma is $\sim 4-10 \ \rm eV$ and HEG resolution is $\sim 70 - 140 \ \rm eV$. Adding thermal broadening has no discernable effect on the predicted counts histogram.

\subsubsection{Accounting for the HETG geometry}\label{sec:hetg-geometry}

\begin{figure*}
    \centering
    \includegraphics[width=0.98\textwidth]{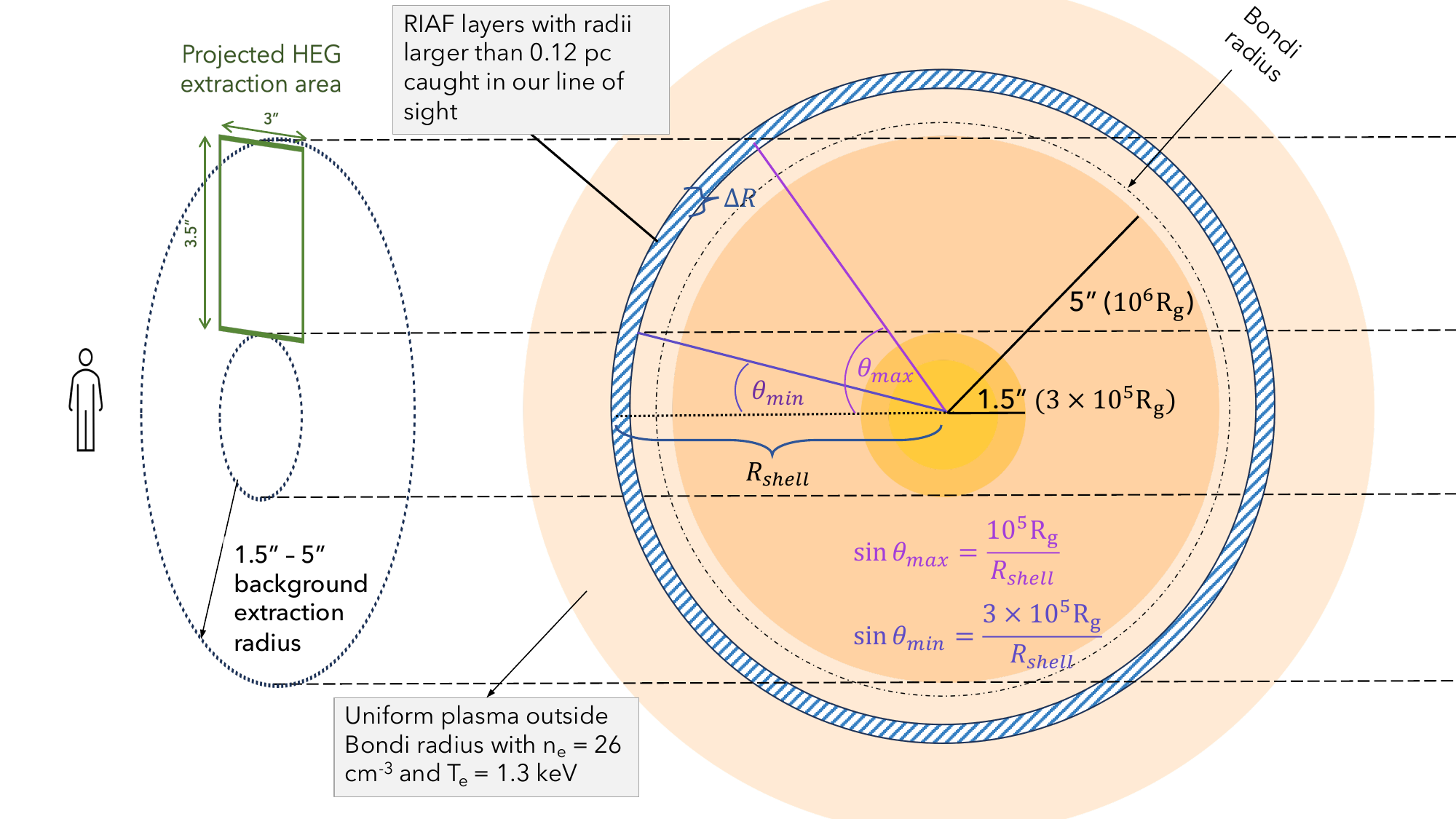}
    \caption{Diagram illustrating the geometry taken into account when calculating the RIAF model flux for the background region. We include the projected HEG extraction area; this reflects how the spectral extraction pipeline for HEG captures photon events between 1.5\arcsec\ and 5\arcsec\ in the cross-dispersion direction and up to 3\arcsec\ along the dispersion axis. In other words, the source spectrum probes the central $3 \times 10^5 R_g$ while the background annulus spans $3 - 10 \times 10^5 R_g$. Only a portion of shells outside of $10^6 R_g$ lie within our line-of-sight projection in the background extraction region (green box). We use $\theta_{min}$ and $\theta_{max}$ to denote the angular boundaries between which a given shell is captured. These quantities are used to calculate the differential volume in each shell, as seen in Equations \ref{eq:dVshell_src} and \ref{eq:dVshell_bkg}. }
    \label{fig:riaf_geometry}
\end{figure*}

The source and background spectra used in this work probe regions of the accretion flow corresponding to $0 - 1.5\arcsec$ ($< 3 \times 10^5 R_g$, or $< 0.06$~pc) and $1.5 - 5\arcsec$ ($3 - 10 \times 10^5 R_g$, or $0.06 - 0.19$~pc) in radius away from \SgrA\, respectively.  To account for layers outside the physical boundaries of the accretion flow that are not completely encompassed by the extraction region but contribute to the plasma emission along the line-of-sight, we integrate the luminosity function along a cylinder, as demonstrated in Figure \ref{fig:riaf_geometry}. We calculate the radial-dependence of the luminosity function, for every shell, as follows:
\begin{equation} \label{eq:dLshell}
\begin{split}
    \frac{d L}{d R} &= n^2 \Lambda(n_e, T_e, Z) \ \frac{d V}{dR} \\
    &= n^2 \Lambda(n_e, T_e, Z) \ 2 R^2~sin\theta ~ d\phi d\theta \\
    &= n^2 \Lambda(n_e, T_e, Z) \ 2 R^2 \times 2 \pi ~ \left[ cos(\theta_{min}) - cos(\theta_{max}) \right] \\
    &= n^2 \Lambda(n_e, T_e, Z)\ 4 \pi R^2 \left( \cos\theta_{min} - \cos\theta_{max} \right)
\end{split}
\end{equation}
where $\Lambda(n_e, T_e, Z)$ comes from \texttt{pyatomdb}, and the differential volume in each shell ($dV$) takes into account our geometric setup, shown schematically in Figure \ref{fig:riaf_geometry}.
Each shell outside the radius of interest (1.5\arcsec\ for the source regions and 5\arcsec\ for the background regions) is multiplied by a geometric factor which takes into consideration the portions of the shell falling within the angular boundaries defined by $\theta_{min}$ and $\theta_{max}$. 
The absolute limits of $\theta$ are $[0,\pi/2]$, and a factor of 2 is introduced into the calculation to account for the emission from both the front and back side of the optically thin accretion flow. 

Another factor to consider is that the HETG order-sorting pipeline only retains events within 3\arcsec\ along the dispersion axis (CPOG Section 8.2), marked by the green box in Figure \ref{fig:riaf_geometry}. For the background model, we include a factor of $f_B = 0.15$, which expresses the ratio of the area captured by the HETG to the area of the 1.5\arcsec$-$5\arcsec\ background extraction annulus. Plugging in the appropriate values for $\theta_{min}$ and $\theta_{max}$ for the various cases, we obtain:
\begin{eqnarray}\label{eq:dVshell_src}
  \left(\frac{d V_{shell}}{dR}\right)_{src} = 4 \pi R^2 \times \hspace{0.6in} \nonumber \\
    \begin{cases}
      1 
      & R < 3 \times 10^5 R_g\\
      \left( 1 - \cos{\theta_{max}} \right) & 3 \times 10^5 R_g < R < 10^6 R_g \\
    \end{cases}
\end{eqnarray}
for the plasma inside the source extraction region, and
\begin{eqnarray}\label{eq:dVshell_bkg}
  \left(\frac{d V_{shell}}{dR}\right)_{bkg} = 4 \pi R^2 f_B \times \hspace{0.6in} \nonumber \\
    \begin{cases}
      1
      & 3 \times 10^5 R_g < R < 10^6 R_g \\
    \left( \cos{\theta_{min}} - \cos{\theta_{max}} \right) & 10^6 R_g < R < 2.4 \times 10^6 R_g \\
    \end{cases}
\end{eqnarray}
for plasma inside the background extraction region. All notation is consistent with that defined in Figure \ref{fig:riaf_geometry}. For both the background and source regions, we integrate out to a maximum radius of $2.4 \times 10^6 R_g$, to ensure appropriate comparison to the simulations in Paper II. We then integrate $dL/dR$ over all shells to get the expected flux arising for a given RIAF with three free parameters: $s$, $q$, $Z_{Fe}$. We treat any material outside the Bondi radius ($1.1 \times 10^6 R_g \approx 5.5\arcsec $) as a uniform plasma with temperature and density $1.3$ keV and $26~\rm cm^{-3}$ \citep{Baganoff2003}.

\subsubsection{Choosing RIAF Model Hyperparameters}\label{sec:model_params}

Several model parameters demanded careful consideration: the metallicity of the plasma, $Z$, the Bondi radius, $R_{Bondi}$, the number of radial bins used in the integration, $N_R$, the innermost radius of integration, $R_{in}$, and the outer bound on the radius, $R_{out}$.

Utilizing \texttt{pyatomdb} grants us significant control over the plasma element abundances. To develop initial estimates of the abundance, we extracted a CCD spectrum from the zeroth-order HETG-S image, following spatial regions used in W13. We used the XSPEC \texttt{tbvarabs} interstellar absorption model \citep{Wilms2000} and incorporated the extinction effect of dust scattering removing light from the source extraction aperture assuming the Rayleigh-Gans approximation, i.e., $ \tau_{sca} = 0.5({\rm N}_{\rm  H}/10^{22}~{\rm cm}^{-2}) (E/{\rm keV})^{-2} $, which is valid for $E > 2$~keV \citep{Predehl1995, Nowak2012, Corrales2016}. The prevalent Iron line at 6.7 keV provides the strongest constraint on the plasma models, and it is poorly fit with standard solar abundances. We obtain a reasonable fit (a reduced Cash statistic of 1.06) with a \texttt{vapec} model where all metal abundances are set to $2~Z_\odot$ \citep[consistent with metallicity measurements from stars in the GC;][]{do2018} and we allow the Iron abundance to vary. While \citet{do2018} measured metallicity in late-type giants, the abundance, and particularly the Iron abundance, in the Wolf-Rayet winds feeding the black hole could be different. The \texttt{vapec} model best fit gives ${\rm N}_{\rm H} = 9.2_{-1.1}^{+0.5} \times 10^{22}~{\rm cm}^{-2}$, $kT = 3.4_{-0.3}^{+0.7}$~keV, and $Z({\rm Fe}) = 1.1 \pm 0.2 Z_\odot$. This relative under-abundance of Iron is consistent with previous results arguing for Iron depletion in the GC ISM \citep{Ponti2016}. Therefore, in our RIAF model procedure, we fix the general plasma abundance to $2~Z_\odot$\citep[with the catalog from GA89;][]{GA89} and only allow the Iron abundance, $Z_{Fe}$, to vary. When calculating the predicted HEG spectrum, we note that the resulting 6.4 keV$-$7.2 keV spectra calculated with $4~Z_\odot$ vs. $2~Z_\odot$ only differ by 0.1\%. 

The canonical Bondi radius for a spherical, non-rotating cloud of gas is determined by the sound speed, $R_{Bondi} = \frac{G M_*}{c_s^2}$, where the sound speed of the plasma is dependent on the metallicity in the form of the mean molecular weight, $\mu$. When incorporating the galactic potential, the Bondi radius has been estimated to be 0.07 pc, or $3.5 \times 10^5 R_g$ \citep{Quataert2002}, while incorporating the rotation of the plasma yields a wider range of values  \citep{Yuan2014}. $R_{Bondi}$ is important in our model because it is the scale factor by which we determined the electron density and temperature at a given radius (see Equation \ref{eq:RIAF}). The canonical values we are using for $n_0$ and $T_0$ ($26$ \ cm$^{-3}$ and $1.3$ keV, respectively) come from \citet{Baganoff2003}, which assumes a Bondi radius of $1\arcsec - 2\arcsec$ and metallicity of $2~Z_\odot$. With $Z \sim 2~Z_\odot$ and an updated mass for \SgrA, the Bondi radius is $R_{Bondi} = 1.1 \times 10^6 R_g$, corresponding to 5.5\arcsec. We note that for a given RIAF $s, q, \text{ and } Z_{Fe}$, changing the value of $R_{Bondi}$ to 4\arcsec\ has no discernable effect on the predicted spectrum.

We find that our RIAF emission model is insensitive to the number of radial bins used, $NR$. We choose to use 150 log-spaced bins from $10^3 - 2.4 \times 10^6 R_g$ over which to calculate the plasma model. 

We chose an inner radius of $R_{in} = 1000~R_g$ because we cannot resolve the inner regions ($1000~R_g \sim 5\times 10^{-3} \arcsec$, while the HETGS angular scale is 0.5731\arcsec). Setting an inner boundary allows us to ignore effects from synchrotron and inverse Compton scattering that arise in the inner regions of the black hole. Close to the black hole's event horizon, magnetic effects start to dominate; many magneto-hydrodynamic simulations of \SgrA\ choose 100 $R_g$ as their outer boundary \citep[e.g.][]{Dexter2020, Ressler2020MHD, Ressler2020GRMHD}.  Outside of $\sim 10^2 - 10^3 R_g$, the electron temperature is too low for significant synchrotron emission \citep{Yuan2003} and significant Comptonization of electrons \citep{Melia1992}. In addition, as we are focusing on a small energy range, any synchrotron or Comptonization effects would add to the continuum and manifest as a powerlaw, which we are already introducing in our fitting procedure (Section~\ref{sec:predicted_spectra}). We note that changing the inner radius from $R_{in} = 1000 R_g$ to $R_{in} = 10^4 R_g$ changes the spectrum by less than 5\%. 

Finally, we consider the outer boundary for the integral described in Section~\ref{sec:hetg-geometry}. The outer limit of the cylindrical projection is important as it determines the contribution from cooler RIAF layers in the model. To maintain consistency with companion paper II, we chose a fixed $R_{out}$ corresponding to 12\arcsec, or $2.4 \times 10^6 R_g$. 

\subsection{Modeling the HEG Spectrum}\label{sec:predicted_spectra}
The RIAF emission must be processed to mimic observation through the HEG. In order to directly compare these models to the extracted HEG spectrum, it is necessary to include effects of interstellar extinction and contaminating background, before calculating the predicted photon counts via folding through the detector response function.

As mentioned in Section \ref{sec:data}, \Chandra\ HETG is a slitless spectrograph. The background stems from both physical plasma and non-physical zeroth order (non-dispersed) image photons, impacting the spectrum. Given our focus on a narrow energy range (6.4 - 7.2 keV), we utilize a power law model to characterize these contaminating photons.
\begin{equation}
    PL = N_{PL} \left( \frac{E}{6.7~\rm keV} \right)^{-\gamma} 
\end{equation} 
and fit for $N_{PL}$ and $\gamma$, where $N_{PL}$ represents the photon flux at 6.7 keV, and $\gamma$ determines the slope of the added powerlaw. 

We modeled the effect of interstellar extinction with the \citet{VY1995} absorption cross-sections for gas-phase elements, following the ISM abundance table of \citet{Wilms2000}, and the dust extinction cross-sections from the \texttt{ISMdust} model \citep{Corrales2016}. We fix the GC column density to $N_H = 10^{23}$~cm$^{-2}$ \citep{Baganoff2003}.
The high column density leads to significant extinction; at 6.7 keV,  we lose 23.2 \% of the photons to gas and dust (8\% to gas absorption, 15.3 \% to dust scattering) along the line-of-sight.

In summary, our final flux model can be described as:
\begin{equation}
\begin{split}
    F^{\rm src}_{HEG} &= F^{\rm src}_{\rm RIAF} (s, q, Z_{Fe}) \ e^{-\tau} + PL(N_{PL}, \gamma) \\
    F^{\rm bkg}_{HEG} &= F^{\rm bkg}_{\rm RIAF} (s, q, Z_{Fe}) \ e^{-\tau} + f_{\rm BACKSCAL}\ PL(N_{PL}, \gamma) 
\end{split}
\end{equation}
where $F_{\rm RIAF}$ is the RIAF model computed from Equation~\ref{eq:dLshell}, integrated over $R$ and divided by $4 \pi d_{\rm \SgrA}^2$, and  $f_{\rm BACKSCAL}$ conveys the scale factor between the background and source extraction region sizes, stored in the `BACKSCAL' data header keyword. We use the publicly available Python package \texttt{pyxsis}\footnote{https://github.com/eblur/pyxsis} to convolve this model flux with the instrumental response. Applying the HEG response function accounts for the energy-dependent effective area and line spread functions of the detector. 
\begin{figure}[h!]
    \centering
    \includegraphics[width=0.48\textwidth]{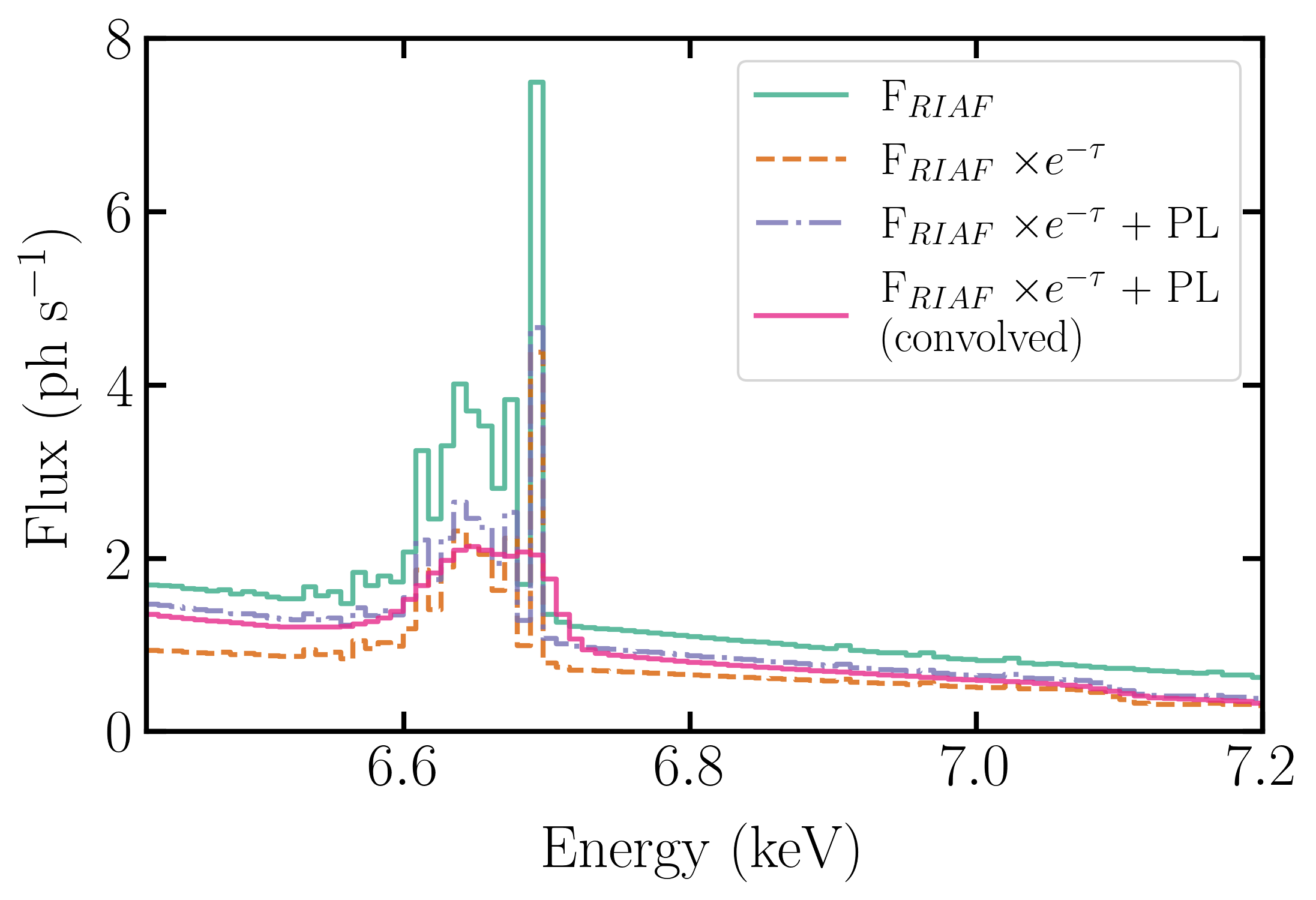}
    \caption{Demonstration of the procedure used to predict the HEG ~+1 source region counts histogram for a given RIAF model. The green curve shows the RIAF emission (in ph s$^{-1}$) calculated with \texttt{pyatomdb}, integrated over the differential volume corresponding to the physical region probed. We incorporate extinction (orange dashed) and a powerlaw for contaminating photons (blue dot-dashed). Convolving with the HEG ancillary response file (ARF) and response matrix file (RMF) leads to the final counts histogram (pink), washing out the fine structure from the original line profiles.}
    \label{fig:procedure}
\end{figure}

In Figure \ref{fig:procedure}, we demonstrate how each step in our processing affects the model spectrum for the HEG~+1 source region. We begin by calculating the RIAF emission (green line) for a given $s, q, \text{and}~ Z_{Fe}$, calculated with the HEG~+1 energy bins and integrated over the effective physical region contributing the spectrum (Figure \ref{fig:riaf_geometry}). Subsequently, we incorporate extinction (orange dashed line) and a powerlaw for contaminating photons (purple dot-dashed line). Lastly, we apply the instrumental  response files to generate the predicted counts histogram (pink). It is clear that the instrumental line spread function washes out much of the fine line structure between 6.6 and 6.7 keV.

\subsection{Fitting}\label{sec:fitting}

We fit the data between 6.4 keV and 7.2 keV using the Python package \texttt{emcee}, which implements Markov chain Monte Carlo (MCMC) sampling with the Metropolis-Hastings algorithm \citep{emcee}. We fit the unbinned background and source region spectra from the HEG +1 and -1 orders simultaneously. In the RIAF models, we fit for $s$ (slope of the electron density distribution), $q$ (slope of the temperature distribution), $Z_{Fe}$ (Iron abundance in units of Z$_{\odot}$), $N_{PL}$ (normalization of the background powerlaw), and $\gamma$ (powerlaw index). The powerlaw component, representing the background, is allowed to be different for the +1 and -1 HEG orders. This gives us in total 7 free model parameters: the $s$, $q$, and $Z_{\rm Fe}$ values, which are the same across both +1 and -1 orders, and four background parameters, $N_{PL}$ and $\gamma$ for each +1 and -1 order.
Due to the small number of counts in each histogram bin, we need to utilize Poissonian statistics. We therefore use the Cash statistic to describe the model likelihood,
\begin{equation}
    C = 2 \Sigma_i \left( m_i - x_i \ln m_i \right)
\end{equation}
where $m_i$ denotes the given model and $x_i$ refers to the data, and $i$ represents each energy bin in the counts histogram. To maximize the likelihood, we minimize the sum of four Cash statistics corresponding to a specified set of model parameters, aggregating $C$ values across both source and background regions for both spectral orders.

\begin{table*}
    \centering
    \begin{tabular}{c|c|c|ll}
 Parameter& \multicolumn{2}{c}{Priors}& \multicolumn{2}{c}{Fit Results}\\  \hline
         &  Gaussian ($\bar{x}, \sigma$)& Limits & Best& (90\% C.I.)\\ \hline 
         s&  - & $ [0,5)$ & 0.74 & $(0.62, 1.40)$\\ \hline 
         q&  - & $ (0,5)$  & 0.22 & $(0.005, 1.56)$\\ \hline 
         Z$_{Fe}$ & (1,2) & (0.001, 10)  & 0.13 & $ (0.03, 0.32)$\\ \hline 
         $\rm \log{N_{PL, +1}}$ & (-6, 3)  & (-12,-3)   & -7.81 & $ (-9.96, -7.53)  $ \\ \hline 
         $\log$N$_{PL, -1}$& (-6, 3)  & (-12,-3)   & -7.29 &$  (-7.52, -7.21) $ \\ \hline 
         $\gamma_{PL, +1}$& - & $ (-20,20) $ & 10.54 & $ (-6.26, 20)  $\\ 
         \hline
         $\gamma_{PL}$& - & $ (-20,20) $ & 3.91 & $  (-0.53, 11.30) $\\ 
         \hline
    \end{tabular}
    \caption{Summary of priors and fit results from our MCMC sampling. The Iron abundance, $Z_{Fe}$ is given in units of $Z_\odot$. $Z_{Fe}$ and the normalization on the powerlaw, $N_{PL}$, are given gaussian priors in additional to hard limits. For each parameter, we list the best-fit value and the 90\% credible intervals.}
    \label{tab:params}
\end{table*}
%

The walkers were guided using the priors detailed in Table \ref{tab:params}. The MCMC was run until  the auto-correlation time $\tau_f$, indicating the number of steps required for the chain to achieve a state independent of its previous state, changed by less than 10\%. We ran the MCMC sampler with 40 walkers for a total of 1600 steps. We excluded the initial 100 steps, after visually inspecting the walkers' exploration of the parameter space, as part of the burn-in phase. This left us with a total of  about $57,000$ samples in the posterior distribution. Table 1 gives the best fit values for each of the seven free model parameters and their corresponding 90\% credible intervals.

\section{Results} \label{sec:results}

\subsection{HETG Spectral Fitting}
\begin{figure*}[h!]
    \centering
    \includegraphics[width=0.48\textwidth]{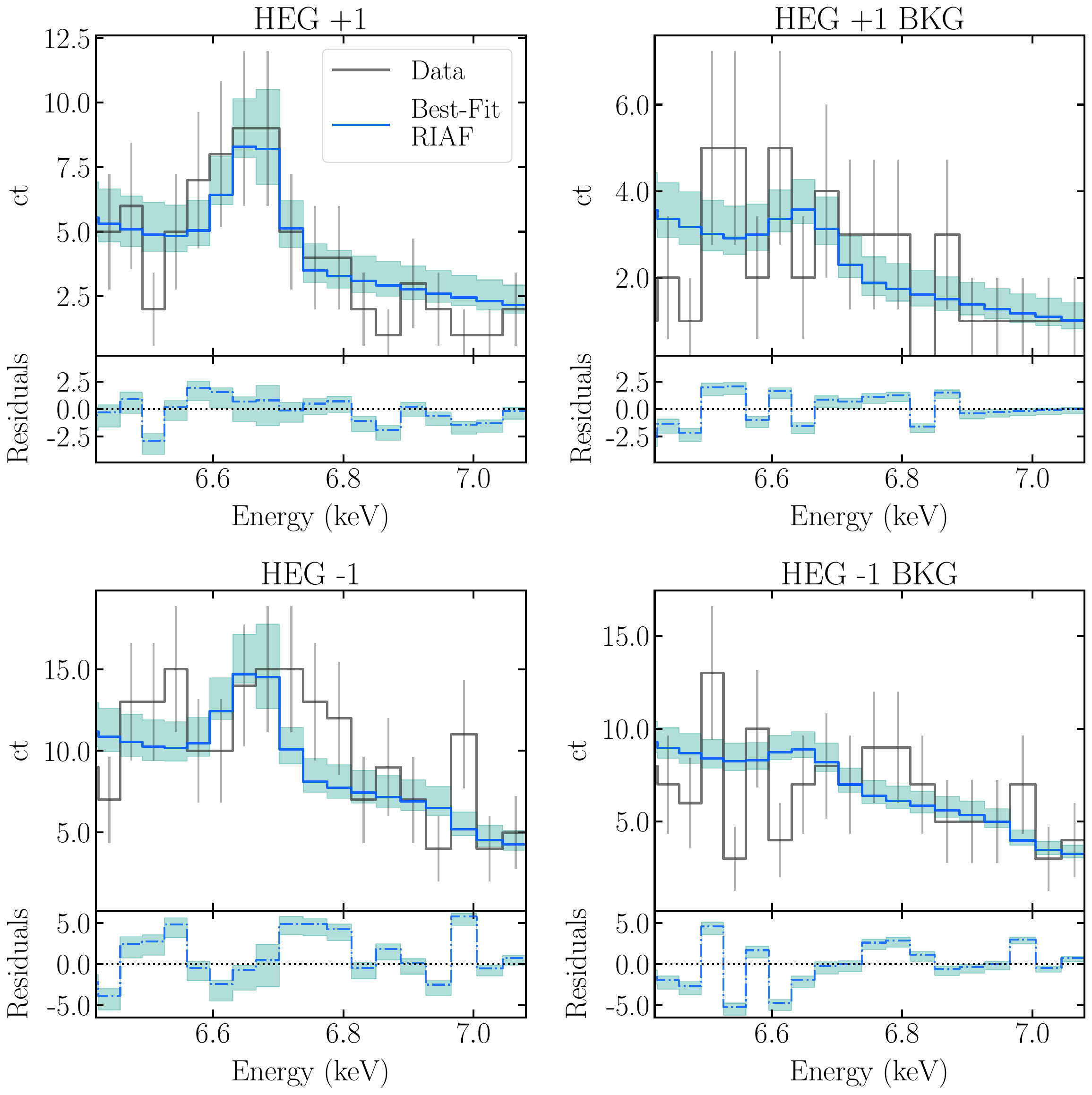}
    \caption{Final fits to the HEG +1 and -1 source and background regions, with error contours corresponding to RIAFs with parameters from  chains within the top 5\% likelihood range. The data and best-fit RIAF are plotted in black and blue, respectively, and are binned to have 4 cts/bin for visual clarity. The residual counts $(x_i - m_i)$ are plotted in dashed blue lines (note the limits on the y-axis are different for the HEG~+1 and HEG~-1 residuals). The model captures some of the 6.7 keV emission but does not capture any structure in the background. }
    \label{fig:finalfit}
\end{figure*}

In Figure \ref{fig:finalfit}, we show the best-fit models determined through our fitting procedure in blue, corresponding to $s = 0.74~(-0.22, +0.66)$, $q = 0.22~(-0.22, +1.35)$, and $Z_{Fe} = 0.13 Z_\odot \ (-0.10, +0.19)$. The best-fit values and 90\% credible intervals for all the model parameters are listed in Table \ref{tab:params}. For visual clarity, the spectra are binned by a factor of 4, and accompanying each spectrum are residual counts, $(x_i - m_i)$.  While the RIAF model effectively captures the structure in the HEG~+1 source region, it predicts a narrower 6.7 keV feature compared to observations from the HEG~-1. Additionally, apparent features in the background spectrum around $6.5$ and $7$ keV are not accounted for by the RIAF model. The HEG~-1 falls on a back-illuminated chip, which has a higher background level, likely explaining the large variance in the HEG~-1 order data. In background regions, the best-fit RIAF model predicts a weak 6.7 keV line that is not observed in the data; here, the power law dominates, providing roughly 95\% of the flux, indicating that most of these photons do not originate from the accretion flow. Due to degeneracies in the best-fit parameter space, predicted counts histograms were constructed using combinations of $s$, $q$, and $Z_{Fe}$ from walkers with the top 5\% of likelihood values. These are used to draw the model contours on top of the best-fitting RIAF models, and it is clear that different combinations of RIAF parameters are difficult to distinguish given the signal in the data.
The number of counts in each bin are too small to make comments on the statistical significance of the residuals in this spectrum.

\begin{figure}[h!]
    \centering
    \includegraphics[width=0.46\textwidth]{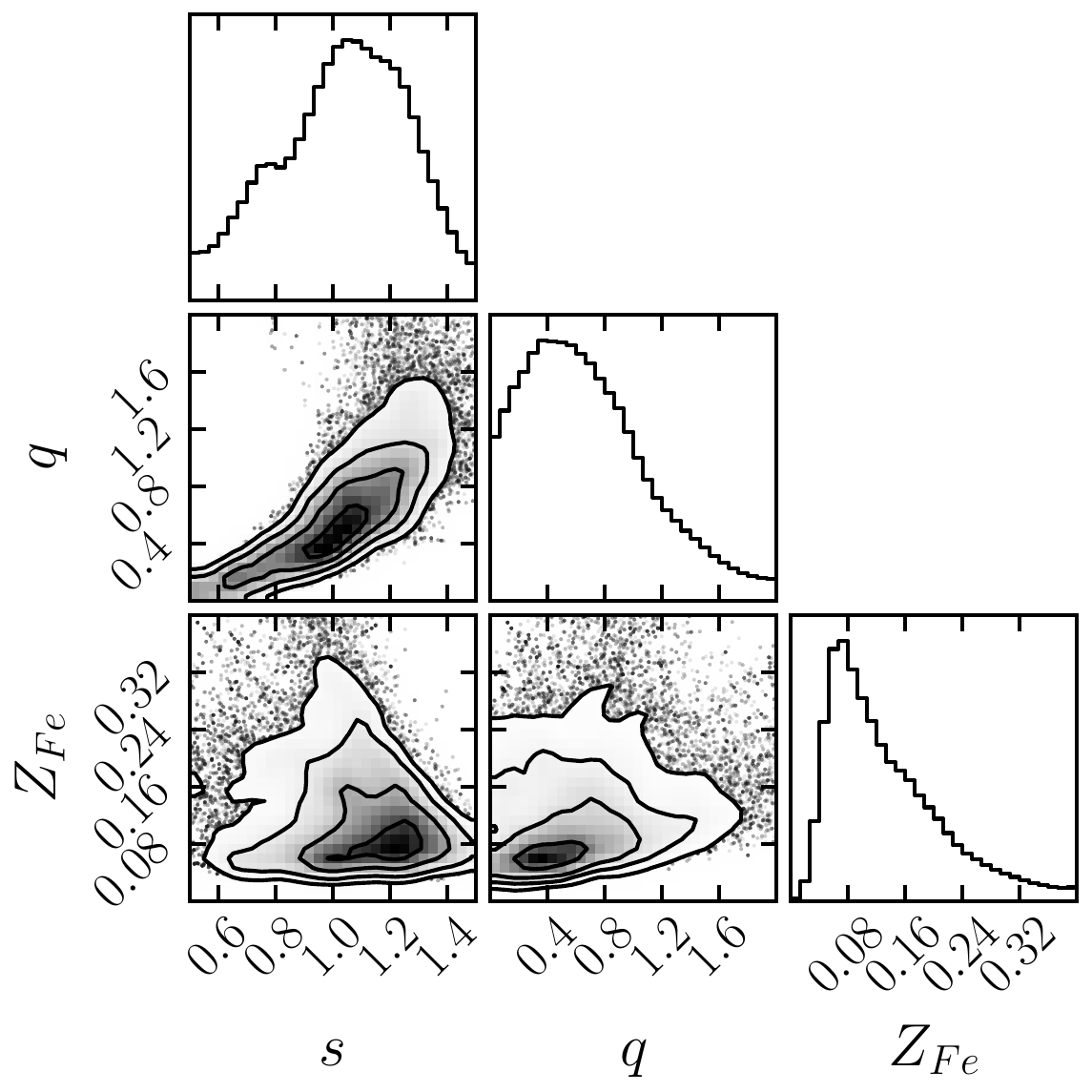}
    
    \caption{LEFT: Corner plot from the final fits of our RIAF model with $\texttt{pyatomdb}$, showing the distributions for RIAF $s$ (material density slope), $q$ (temperature profile slope), and $Z_{Fe}$ (relative to solar; all other elemental abundances were set to $2\ Z_\odot$). Overall, the data prefer sub-solar Iron abundances, $s$ values close to 1, and $q \approx 0.22$. }
    \label{fig:corner}
\end{figure}

Figure \ref{fig:corner} shows the posterior distribution obtained for $s$, $q$, and $Z_{Fe}$. Among these parameters, we observe that the slope of the temperature profile, $q$, and slope of the density profile, $s$, are degenerate. However, the distributions for each parameter are quite well-defined and constrained. The best fit suggests that $s = 0.74$, indicating an inflow slightly stronger than an outflow in this region. While this aligns with the findings of W13, the best-fit temperature profile has $q = 0.22$, indicating a much shallower profile than the assumed $q \sim 1$ in W13. According to conservation of energy arguments within a gravitational potential where $U = \frac{1}{r}$, the temperature profile should theoretically show a linear decrease with radius \citep[][]{Quataert2002}. Such a flat temperature profile has been commonly assumed in ADAF models \citep{Begelman2012, Yuan2014}. In contrast, our RIAF model predicts a less steep dependence on temperature. The data demonstrably prefer low values of $Z_{Fe}$, indicating a subsolar Iron abundance that is broadly consistent with the results of the APEC fits (Section \ref{sec:model_params}) and \citet{Ponti2016}.

\section{Summary \& Conclusions} \label{sec:conclusions}

As the closest SMBH, low-luminosity \SgrA\ offers unique opportunities to study inefficient accretion flows. The quiescent high-resolution \Chandra\ X-ray Visionary Program (PIs: Markoff, Nowak, \& Baganoff)\footnote{https://www.sgra-star.com/} spectrum of \SgrA\ contains line structure that could hint at potential temperature structure, but could also be an artifact of extraction techniques and noise \citep{Corrales2020}. To better understand this spectrum, we develop a model for emission via the RIAF prescription \citep[e.g.][]{Yuan2014}, accounting for the specific geometry and extraction procedures of the \Chandra\ HETG-S instrument. RIAFs are characterized by power-law density ($n \propto r^{-3/2 + s}$) and temperature ($T \propto r^{-q}$) profiles, with $s = 0$ and $s = 1$ corresponding to classical Bondi accretion, where mass is conserved in the accretion inflow, and a balanced inflow and outflow, respectively. 

We calculate the RIAF emission under the assumption of collisional ionization equilibrium within nested spherical shells. The \texttt{pyAtomDB} code is employed for photon emissivity modeling in each shell, considering optically thin, spherically symmetric plasma in the region that is physically probed by various HETG-S extraction regions. We choose a general metal abundance of $2~Z_\odot$ to model the plasma \citep{do2018}, but allow the Iron abundance, $Z_{Fe}$, to vary. We note that our RIAF model depends on certain hyperparameters which change the resulting counts histogram by less than 10\%, a difference that cannot be resolved with the current dataset.

We simultaneously fit HEG~+1 and HEG~-1 source and background regions with our RIAF emission model. Our model is unable to capture any line features other than the 6.7 keV He-like Iron lines, implying that a majority of the line structure arises due to a high X-ray background in the region and statistical noise. The elevated X-ray background indicates that the photons we detect outside of 1.5\arcsec\ may not come from the \SgrA\ accretion flow but rather from numerous unresolved point sources within this region \citep{Muno2003}.

When allowing both $s$ and $q$ to vary, we find a best fit of $s = 0.74_{-0.12}^{+0.66}$, indicating an inflow and outflow balance consistent with both previous fits to the non-dispersed \Chandra\ X-ray spectrum (W13) and multiwavelength models with boundary conditions fixed to the accretion rate at the Bondi radius \citep[$s \approx 0.6$][]{Ma2019}. The posterior distribution we obtained for $q$ suggests a shallower temperature profile than that typically assumed by ADAF models \citep{Quataert2002, Yuan2003, Begelman2012}. Our best fit is $q = 0.22_{-0.22}^{+1.35}$. The data also require a low Iron abundance, $Z_{Fe} < 0.32 Z_\odot$. Considering measurements of stars in the GC \citep[$Z \sim 2~Z_\odot$;][]{do2018}, this argues for Iron depletion in the GC, and is consistent with \citet{Ponti2016}. The slopes of the density and temperature profiles are degenerate with each other. In a companion paper (Paper II, Balakrishnan et al. 2024b), we compare the RIAF model to predictions of the shocked plasma arising from colliding winds in the central parsec \citep{Cuadra2015, Russell2017}, and provide microcalorimeter predictions for both these models.

\begin{acknowledgments}
We would like to thank the anonymous referee for their helpful and insightful responses. The original \Chandra\ dataset used in this work was made possible by the \Chandra\ X-ray Visionary Program through \Chandra\ Award Number GO2-13110A, issued by the \Chandra\ X-ray Center (CXC), which is operated by the Smithsonian Astrophysical Observatory for NASA under contract NAS8-03060. Authors MB, LC, and CR were supported by the CXC grants program, award AR2-23015A and AR2-23015B. DC is funded by the Deutsche Forschungsgemeinschaft (DFG, German Research Foundation) under Germany’s Excellence Strategy – EXC 2121 – ‘Quantum Universe’ - 390833306. JC acknowledges financial support from ANID (FONDECYT 1211429 and Millenium Nucleus TITANS, NCN2023\_002). DH acknowledges funding from the Natural Sciences and Engineering Research Council of Canada (NSERC) and the Canada Research Chairs (CRC) program. 
\end{acknowledgments}

\software{matplotlib \citep{Hunter:2007}, astropy \citep{2013A&A...558A..33A,2018AJ....156..123A}, pyatomdb \citep{pyatomdb}, pyxsis \footnote{https://github.com/eblur/pyxsis}
          }




\bibliography{refs}{}
\bibliographystyle{aasjournal}



\end{document}